\documentclass[%
reprint,
 amsmath,amssymb,
 aps,
]{revtex4-2}

\usepackage{graphicx}
\usepackage{bm}

\usepackage[utf8]{inputenc}
\usepackage[T1]{fontenc}
\usepackage{mathptmx}
\usepackage{braket}
\usepackage{float}
\usepackage{tasks}
\usepackage{xcolor}
\usepackage{dblfloatfix}
\usepackage{array}
\usepackage{makecell}
\usepackage{amsmath}
\usepackage{gensymb}
\usepackage[caption=false]{subfig}
\usepackage{hyperref}

\begin{document}

\preprint{APS/123-QED}

\title{Polarization-orbital angular momentum duality assisted entanglement observation for indistinguishable photons}

\author{Nijil Lal}
\email{nijil@prl.res.in}
\affiliation{Physical Research Laboratory, Ahmedabad 380009, India}

\author{Sarika Mishra}
\affiliation{Physical Research Laboratory, Ahmedabad 380009, India}
\affiliation{Indian Institute of Technology, Gandhinagar 382355, India}

\author{Anju Rani}
\affiliation{Physical Research Laboratory, Ahmedabad 380009, India}
\affiliation{Indian Institute of Technology, Gandhinagar 382355, India}

\author{Anindya Banerji}
\affiliation{Physical Research Laboratory, Ahmedabad 380009, India}

\author{ Chithrabhanu Perumangattu}
\affiliation{Centre for Quantum Technologies, National University of Singapore, 3 Science Drive 2, S117543, Singapore}

\author{R. P. Singh}
\affiliation{Physical Research Laboratory, Ahmedabad 380009, India}

\date{\today}

\begin{abstract}
Duality in the entanglement of identical particles manifests that entanglement in only one variable can be revealed at a time. We demonstrate this using polarization and orbital angular momentum (OAM) variables of indistinguishable photons generated from parametric down conversion. We show polarization entanglement by sorting photons in even and odd OAM basis, while sorting them in two orthogonal polarization modes reveals the OAM entanglement. The duality assisted observation of entanglement can be used as a verification for the preservation of quantum indistinguishability over communication channels. Indistinguishable photons entangled in complementary variables could also evoke interest in distributed quantum sensing protocols and remote entanglement generation.
\end{abstract}

\maketitle

\section{\label{sec:level1}Introduction}
\maketitle

Complementarity is a unique manifestation of quantum mechanics, like entanglement. Introduced as a concept by Niels Bohr \cite{bohrcompli} and developed through further rigorous scientific discussion, complementarity broadly states that objects possess mutually exclusive properties such that the full knowledge of one property precludes full knowledge of the conjugate one. The wave-particle duality \cite{WZineq} and Heisenberg's uncertainty principle \cite{Kraus} are closely associated with the complementarity principle. \textit{Welcher Weg} experiments demonstrate the complementarity between distinguishability and interference visibility implying that quantum interference will take place only if the measurement does not distinguish between the interfering pathways \cite{greenberger88,qeraser,pittman,whichwayAKpati}. In other words, indistinguishability leads to quantum interference. It is interesting to note that interference is related to coherence, which is a wave property, while distinguishability is associated with localized variables, which is a particle-like property \cite{Mandel91,QDuality}. Indistinguishability of photons evoke great interest in quantum information protocols \cite{indis_1,indis_2}. In fact, many studies have aimed towards reducing the distinguishability in entangled systems \cite{branning,torresindist,killoran2014} to achieve higher visibility of quantum interference.

Indistinguishability requires perfect overlap in spatio-temporal position, energy, polarization etc. which can be demonstrated through two-photon interference experiments. Indistinguishable photons could be generated in the output of a Hong-Ou-Mandel interferometer \cite{hom1987}. A source of highly indistinguishable and entangled photon pairs is crucial to various quantum information applications \cite{sc_indist1,sc_indist2, lo_indist3}. However, it is not possible to observe the entanglement between such indistinguishable photons unless we sort and separate them in terms of a physical variable, such as their position, momentum, polarization, orbital angular momentum (OAM) etc. For a general case of degenerate, non-collinear type-II spontaneous parametric down-conversion (SPDC) output, we can write the identical photons in terms of their different degrees of freedom as,
\begin{equation}
    \left|\Psi\right\rangle=\frac{1}{\sqrt{2}}\left(\left|H, \mathbf{k}_{i}\right\rangle\left|V, \mathbf{k}_{s}\right\rangle+\left|V, \mathbf{k}_{i}\right\rangle\left|H, \mathbf{k}_{s}\right\rangle\right)
\end{equation}
where the state is written in terms of polarization and linear momentum. This can also be written as a polarization entangled state,
\begin{equation}
    \left|\Psi\right\rangle=\frac{1}{\sqrt{2}}\left(\left|H\right\rangle_{k_{i}}\left|V\right\rangle_{k_{s}}+\left|V\right\rangle_{k_{i}}\left|H\right\rangle_{k_{s}}\right)
\end{equation}
using their linear momentum as a label to differentiate the subsystems. It is also possible to express the same state as,
\begin{equation}
    \left|\Psi\right\rangle=\frac{1}{\sqrt{2}}\left(\left|k_{i}\right\rangle_{H}\left|k_{s}\right\rangle_{V}+\left|k_{s}\right\rangle_{H}\left|k_{i}\right\rangle_{V}\right).
\end{equation}
which is entangled in linear momentum and the individual subsystems are labelled by their polarization. This complimentary behaviour of two independent degrees of freedom of identical particles is called duality in entanglement \cite{bose2013duality,bhattigsa,moreva2015bell}. To demonstrate this, we use another degree of freedom, namely orbital angular momentum. Considering a type-II SPDC process, the generated photon pairs are independently entangled in polarization and OAM, following the birefringence properties of the crystal and OAM conservation ($l_{p} = l_{1} + l_{2}$) \cite{mair2001,ali_NJP,nijil_JMO}. The corresponding state can be expressed as,
\begin{equation}
\begin{aligned}
    |\Psi\rangle= &\frac{1}{2}( |H, l_{1}, \boldsymbol{k}_{i}\rangle|V, l_{2}, \boldsymbol{k}_{s}\rangle+|H, l_{2}, \boldsymbol{k}_{i}\rangle|V, l_{1}, \boldsymbol{k}_{s}\rangle \\
    & + |V, l_{1}, \boldsymbol{k}_{i}\rangle|H, l_{2}, \boldsymbol{k}_{s}\rangle+|V, l_{2}, \boldsymbol{k}_{i}\rangle|H, l_{1}, \boldsymbol{k}_{s}\rangle ).
\end{aligned}
\label{eq:klabel}
\end{equation}
For a collinear output, the linear momentum labelling given in Eq. (\ref{eq:klabel}) becomes unavailable since $k_i = k_s = k$ and it becomes impossible to observe the entanglement. However, one can write,
\begin{equation}
\begin{aligned}
    |\Psi\rangle &=\frac{1}{\sqrt{2}}\left(|H\rangle_{l_{1}}|V\rangle_{l_{2}}+|V\rangle_{l_{1}}|H\rangle_{l_{2}}\right) \otimes|\boldsymbol{k}\rangle \\
    &\equiv \frac{1}{\sqrt{2}}\left(|l_{1}\rangle_{H} |l_{2}\rangle_{V}+|l_{2}\rangle_{H} |l_{1}\rangle_{V}\right) \otimes|\boldsymbol{k}\rangle .
\end{aligned}
\end{equation}
In most protocols involving the entanglement of orbital angular momentum of photons, the infinite dimensional OAM spectrum in the output of SPDC is restricted to a two-dimensional basis by the post-selection of the twin-photons. Due to this post-selection, a large amount of generated photons which belong to the other states in the infinite dimensional OAM basis are lost. A method to avoid this loss is to use an alternate basis defined by the even and odd states of OAM \cite{cbhanuevenodd, cbhanuhyper}.

In this work, we propose that the OAM of twisted photons defined in their even-odd basis can be used to separate the otherwise completely indistinguishable photons in the collinear output. For a pump beam carrying an odd OAM value, the SPDC photons will be generated in pairs of even and odd OAM states, following the conservation of OAM. For pump OAM, $l_p = 1$, in a collinear type-II SPDC process where the idler-signal pairs are generated in orthogonal polarization states, the output OAM state can be written as,
\begin{equation}
\begin{aligned}
    |\Psi\rangle_{SPDC} &= \sum_{m=-\infty}^{+\infty} c_{m,1-m}|m\rangle_{H}|1-m\rangle_{V}\\
&= c_{0,1}|0\rangle_{H}|1\rangle_{V}+c_{1,0}|1\rangle_{H}|0\rangle_{V}\\ &\hspace{0.3cm} + c_{2,-1}|2\rangle_{H}|-1\rangle_{V}+c_{-1,2}|-1\rangle_{H}|2\rangle_{V} + ...\\
&= \frac{1}{\sqrt{2}}(|E\rangle_{H}|O\rangle_{V}+|O\rangle_{H}|E\rangle_{V}).
\label{eq:evenoddHV}
\end{aligned}
\end{equation}
Here, the twin photons are indistinguishable in every other degree of freedom, including their spatial position, except for their polarization and OAM. However, it is not possible to make individual measurements on these photons unless we separate them under some label. We use their polarization state to label these individual photons and observe the entanglement in the even-odd basis of the OAM. In the same way, the state in Eq. (\ref{eq:evenoddHV}) can be written by labelling them as even and odd OAM states,
\begin{equation}
    |\Psi\rangle_{SPDC} =  \frac{1}{\sqrt{2}}(|H\rangle_{E}|V\rangle_{O}+|V\rangle_{E}|H\rangle_{O}).
\end{equation}
Hence, depending upon whether the polarization or OAM has been used for the labelling, we can observe the entanglement in the other degree of freedom. It can be seen that the distinguishability of the associated particles reveal the entanglement. Experimentally, this means that the method that we use to distinguish signal and idler photons dictates the degree of freedom in which the photons are entangled.

Sorting of photons based on the polarization can be achieved using a simple polarizing beam splitter which separates H and V, revealing the entanglement in even-odd OAM states. To observe the polarization entanglement, we use an even-odd OAM sorter. The conventional and commonly used method for selecting and measuring the OAM component of photons is the phase-flattening technique. The spiral phase corresponding to a desired OAM state is flattened using spiral phase plates or holograms displayed on spatial light modulators. The resulting fundamental Gaussian mode is then coupled to a single mode fiber for measurement. However, such projective measurements based on phase-flattening have shortcomings in terms of efficiency and dependence on pump characteristics \cite{pflimitns}. Moreover phase flattening projects the photon into one of the OAM states and hence cannot be used in even-odd sorting. Using the two-dimensional even-odd basis for the twin-photon OAM states, the efficiency of entanglement protocols could be increased.

\begin{figure}[h]
\centering
    \subfloat{%
    \includegraphics[width=0.5\columnwidth]{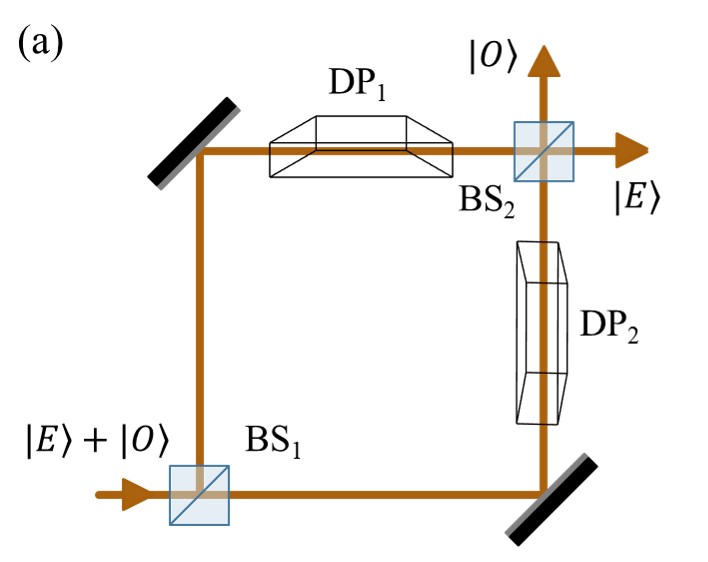}\label{fig:sorter_MZ_a}%
    }
   \subfloat{%
    \includegraphics[width=0.5\columnwidth]{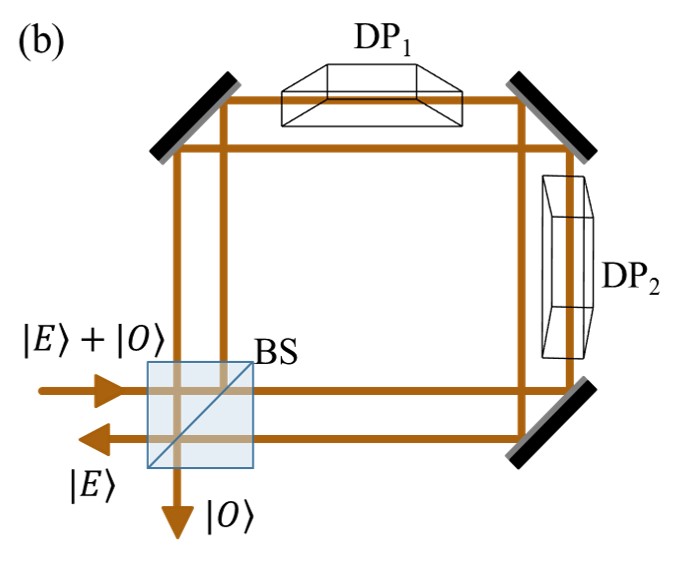}\label{fig:sorter_MZ_b}%
    }
    \caption{Interferometric sorter for even and odd OAM states of light using Dove prism in (a) Mach-Zehnder arrangement and (b) Folded Mach-Zehnder arrangement. In both figures, the orientation of the two Dove prisms are perpendicular to each other.}
        \label{fig:sorter_MZ}
\end{figure}

Figure \ref{fig:sorter_MZ_a} illustrates a basic set up for an even-odd OAM sorter that involves a Mach-Zehnder interferometer with a Dove prism in each arms \cite{leachint,sorterCNOT}. A dove prism flips the OAM from $+l$ to $-l$ during the internal reflection. When two Dove prisms are kept in the two arms of an interferometer, it introduces an OAM dependent relative phase $2l\alpha$ where $\alpha$ is the relative rotation of the Dove prisms. The two Dove prisms are oriented perpendicular to one another and hence $\alpha = \pi/2$. This introduces a phase $l\pi$ between the two arms of the interferoemeter.
The relative phase difference would turn out to be odd multiples of $\pi$ for all odd OAM orders and even multiples of $\pi$ for even OAM orders. As a result, the constructive interference will take place in different output ports for even and odd OAM values. To overcome the stability concerns, the Mach-Zehnder arrangement in Fig. \ref{fig:sorter_MZ_a} could be reconfigured as a more robust polarizing Sagnac interferometer \cite{cbhanu_sagnacsorter,slussarenkosorter}. However, a polarizing interferometer destroys the indistinguishability in polarization. In this experiment, we adopt a folded Mach-Zehnder arrangement [Fig. \ref{fig:sorter_MZ_b}] to set up a robust interferometer without affecting the indistinguishability as well as the duality.

\section{Methodology}

We take a periodically-poled type-II KTP (ppKTP) crystal to observe the duality in entanglement of polarization and OAM of twin photons generated in SPDC. In type-II SPDC, the idler and signal photons will have perpendicular polarizations along ordinary and extra-ordinary axes of the crystal. The output polarization state can be written as,
\begin{equation}
    \Psi = \frac{1}{\sqrt{2}} \ket{H}_i \ket{V}_s \pm \ket{V}_i \ket{H}_s .
\end{equation}
Before going for the collinear indistinguishable photons, we consider the non-collinear generation of photon-pairs where the two photons are emitted along directions following the phase-matching condition,
\begin{equation*}
{\bold{k}_i} + {\bold{k}_s} \approx {\bold{k}_p}
\end{equation*}
where $\bold{k}$ is the linear momentum vector. The down-converted output forms a cone of correlated signal-idler photon pairs following the non-collinear phase matching condition.
\begin{figure}[h]
    \centering
    \includegraphics[width=\columnwidth]{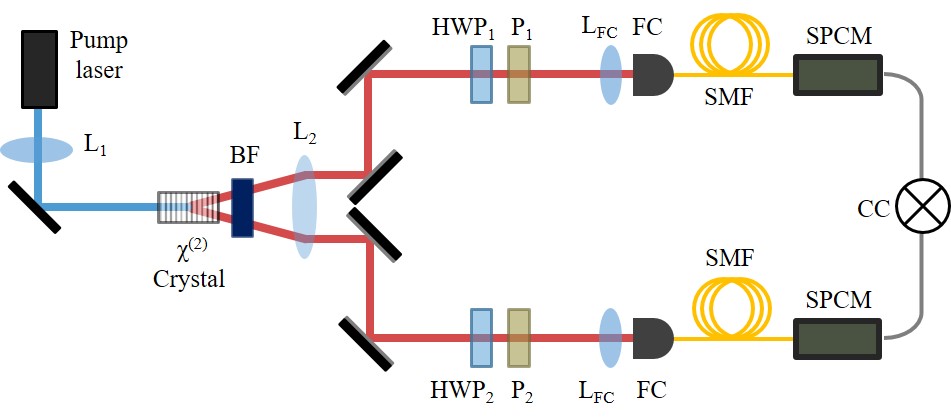}
    \caption{Experimental setup to observe the entanglement in a non-collinear type-II SPDC from a ppKTP crystal ($\chi^{(2)}$). Polarization measurements are carried out using a combination of half-wave plate (HWP) and a polarizer (P) in each arm. $\mathrm{L_{FC}}$ - Aspheric lens associated with the fiber coupler (FC), SMF - Single mode fiber, SPCM - Single photon counting module, CC - Coincidence counter.}
    \label{fig:noncoll_setup}
\end{figure}
The experimental schematic is given in Fig. \ref{fig:noncoll_setup}. We use a Toptica TopMode (405 nm) laser as the pump which is loosely focused within the crystal using the lens, $\mathrm{L_1}$ (f=50 cm) such that the paraxial approximation for OAM conservation is valid. A band-pass filter (BF, 810 $\pm$ 5 nm) blocks the residual pump while transmitting the down-converted output.
\begin{figure}[h]
    \centering
    \includegraphics[width=\columnwidth]{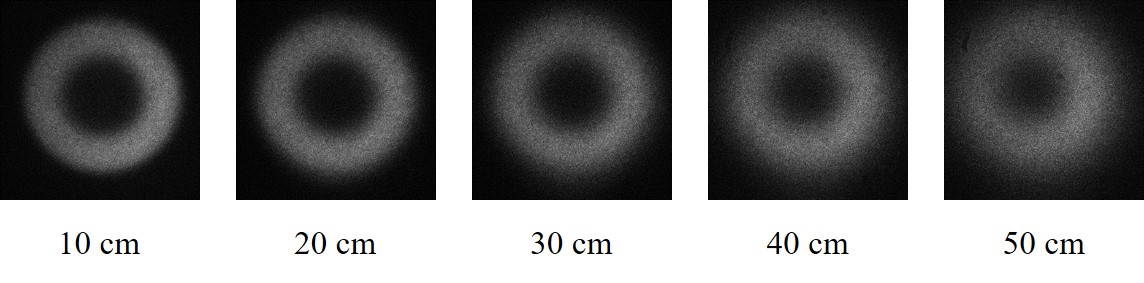}
    \caption{The collimated SPDC output images obtained using an EMCCD kept at different distances from the crystal plane. A 10 cm lens placed after the SPDC output collimates the diverging cone of photon pairs.}
    \label{fig:collimNC}
\end{figure}
With this spectral filtering, we consider our source to be giving degenerate photon pairs in the setups discussed in this work. The diverging cone of SPDC photons is collimated using lens, $\mathrm{L_2}$ (f=10 cm), for the overall length of the experiment [Fig. \ref{fig:collimNC}]. The low intensity photon distributions are imaged using an EMCCD (Andor iXon3). Signal and idler photons are then coupled to single mode fibers (SMF) through fiber coupling systems, FC (Thorlabs CFC-5X-B), consisting of a fiber launcher and an aspherical lens ($\mathrm{L_{FC}}$, f = 4.6 mm). These photons are detected at single photon counting modules, SPCM (Excelitas AQRH-16-FC), whose output is given to a coincidence counter, CC (ID Quantique ID800 TDC), to obtain the coincidence counts. To observe maximum entanglement, maximum overlap between the two polarization modes has to be ensured.
\begin{figure}[h]
\centering
\includegraphics[width=\columnwidth]{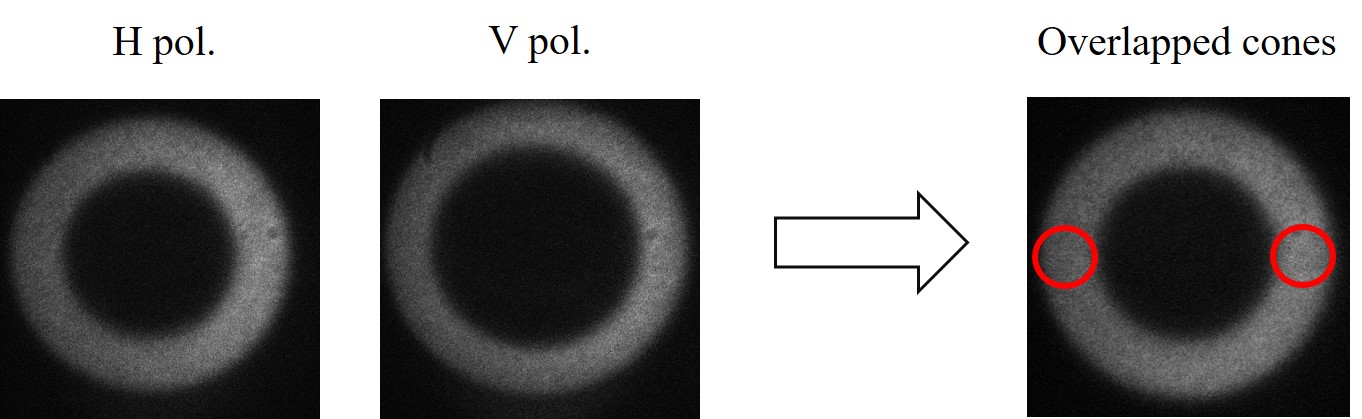}
\caption{The crystal position and tilt are adjusted such that the SPDC output corresponding to both the polarizations are spatially overlapping. Diametrically opposite regions (in red circles) corresponds to the entangled photons.}
\label{fig:OverlapHV}
\end{figure}
The crystal position and tilt for pump incidence are adjusted such that good spatial overlap between the H-polarized cone and V-polarized cone is achieved [Fig. \ref{fig:OverlapHV}]. The correlated photons will be falling along diametrically opposite points following the phase matching conditions. Two regions, marked in red circles, are selected and coupled to the detector system. 

The polarization projections are done using the combination of half wave plate (HWP) and a polarizer (P) kept in each arm of the SPDC output [Fig. \ref{fig:noncoll_setup}].
\begin{figure}[h]
\centering
\includegraphics[width=\columnwidth]{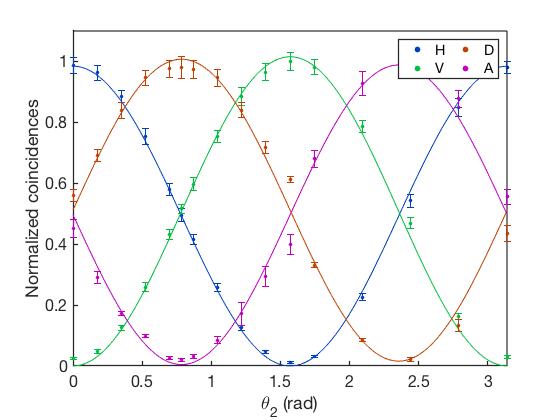}
\caption{Polarization correlations corresponding to the spatially separated photons in the non-collinear down conversion pair. Error bars indicate statistical uncertainty of one standard deviation.}
\label{fig:vis_noncoll}
\end{figure}
The half wave plate, $\mathrm{HWP_1}$, in the idler arm is kept at fixed angles of rotation, $\theta_1$ = 0$\degree$, 45$\degree$, 90$\degree$ and 135$\degree$ and the second half wave plate kept in the signal arm ($\mathrm{HWP_2}$) is rotated as $\theta_2$ to observe the polarization correlation. The experimentally observed polarization correlations, corresponding to the projections in HV basis as well as DA basis, are given as the visibility curves in Fig. \ref{fig:vis_noncoll}. The normalized coincidences are plotted along y-axis with the variation of $\theta_2$ for (blue) $\theta_1$ = 0$\degree$, (red) $\theta_1$ = 45$\degree$, (green) $\theta_1$ = 90$\degree$ and (purple) $\theta_1$ = 135$\degree$ where $\theta_1$ and $\theta_2$ correspond to the rotation of $\mathrm{HWP_1}$ and $\mathrm{HWP_2}$ respectively in Fig. \ref{fig:noncoll_setup}. The observed visibilities in two bases are 96.4$\pm$0.2\% (HV basis) and 94.1$\pm$0.3\% (DA basis) respectively. The Bell parameter is estimated to be, S = 2.69 $\pm$ 0.03.

The non-collinear down-converted photon pairs discussed above are labelled by their spatial positions, and hence distinguishable. A collinear output can be assumed to give indistinguishable photon pairs since they are not separated in spatial modes.
\begin{figure}[h]
    \centering
    \includegraphics[width=\columnwidth]{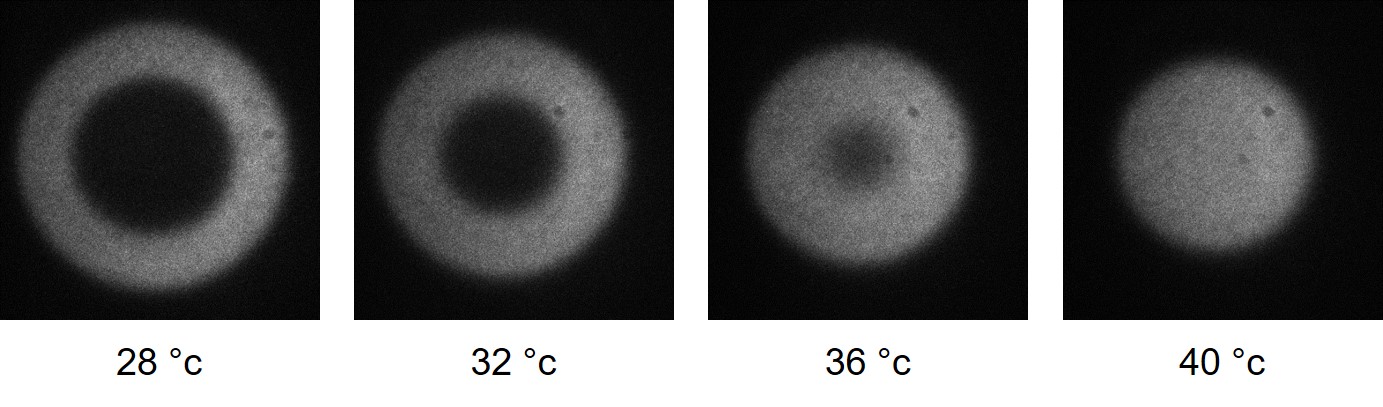}
    \caption{Variation of the spatial distribution of the SPDC output from a type-II ppKTP crystal with different temperature values for quasi-phase matching. We observe that collinear output is obtained for a temperature, 40 $\degree$C.}
    \label{fig:tempPM}
\end{figure}
For a periodically poled crystal, the phase matching will be governed by the temperature of the crystal \cite{ppKTP_1,ppKTP_2} and hence the crystal is mounted on a temperature controlling oven. By varying the temperature, one can achieve collinear phase matching condition as shown in Fig. \ref{fig:tempPM}. For our crystal, the collinear output is obtained at 40 $\degree$C.
\subsection{Observation of polarization entanglement through OAM sorting}
We use a double Mach-Zehnder OAM sorting interferometer to separate the photons on the basis of their OAM and reveal the entanglement in polarization. A double Mach-Zehnder type interferometer could be understood as a normal Mach-Zehnder interferometer, folded back such that the input and output beam splitters become the same. Such a configuration will have the stability of a common path interferometer, since both the arms see same optical components, and the ease of inserting independent components in the interfering arms as in a Mach-Zehnder interferometer.

\begin{figure}[h]
    \centering
    \includegraphics[scale=0.5]{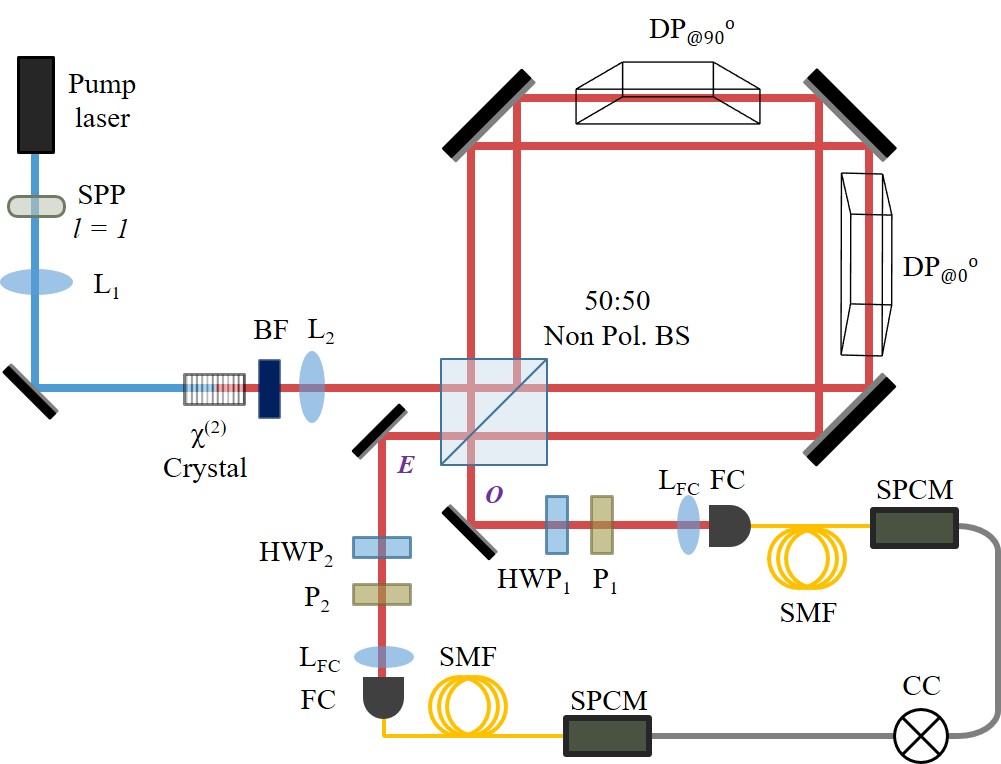}
    \caption{Schematic to sort the even-odd states of OAM from a collinear SPDC with pump carrying OAM ($l_{p}=1$). The Dove prisms within the double Mach-Zehnder interferometer is kept orthogonal to each other. Half wave plate (HWP) along with polarizer (P) corresponds to the polarization projectors. O refers to the constructive port for odd OAM and E labels the constructive port for even OAM.}
    \label{fig:OAMsortPOLent}
\end{figure}

The collinear correlated pairs of photons having even and odd OAM orders are sent to a double Mach-Zehnder interferometer containing two Dove prisms which are kept in the individual paths as given in Fig. \ref{fig:OAMsortPOLent}. The Dove prisms are kept such that their relative orientation is perpendicular to each other. The individual photons undergo single photon interference within the interferometric sorter. Photons carrying an odd OAM will constructively interfere in the odd port (O) whereas photons having even OAM will show up in the even port (E). The polarization projections are done using the combination of a half wave plate and a polarizer kept in each output port.
At first, an alignment laser beam (810 nm, Thorlabs) is used to verify the sorting of even and odd OAM modes. The collinear output is then sent along the same path. Coincidences are maximized in the detectors kept in ports E and O, and polarization projection measurements are carried out to observe the entanglement visibility.
\subsection{Observation of OAM entanglement through polarization sorting}
The entanglement in the even-odd basis of orbital angular momentum can be observed by separating the indistinguishable photons with their polarization as a label.
\begin{figure}[h]
\centering
    \includegraphics[width=0.48\textwidth]{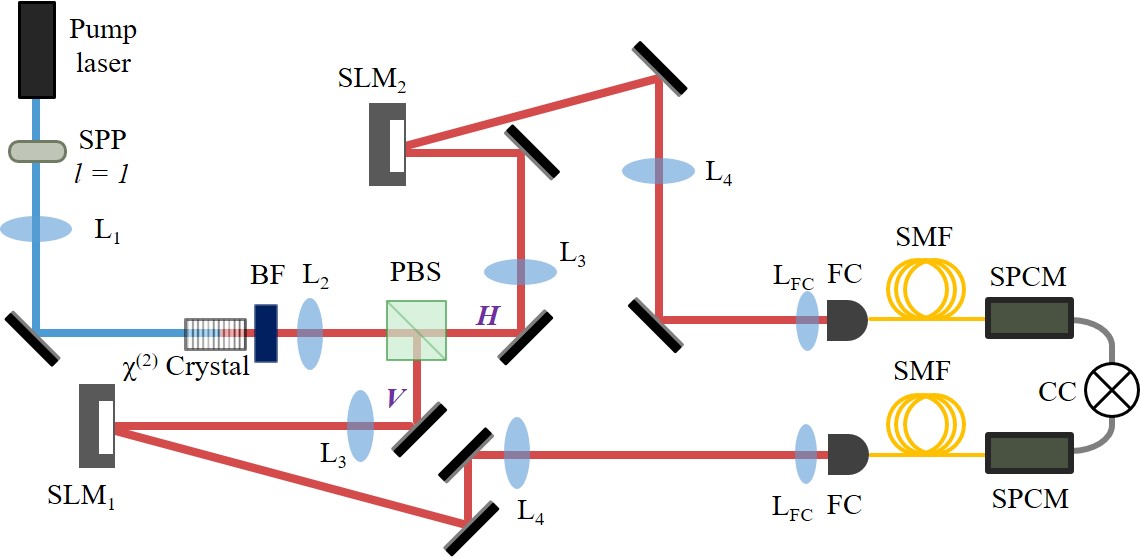}
    \caption{Schematic to observe entanglement in the even-odd basis of OAM by sorting photons in polarization. Indistinguishable photons in the type-II down-converted pairs are sorted in polarization using a polarizing beam splitter (PBS).}
    \label{fig:POLsortOAMent}
\end{figure}
A polarizing beam splitter (PBS) separates the H and V polarized photons in the transmitted and reflected ports respectively as given in Fig. \ref{fig:POLsortOAMent}. OAM projections are done using a spatial light modulator (SLM) through phase-flattening and coupling to single mode fibers. Additional lenses are used to image the crystal plane to the SLM, as well as the SLM plane to the coupling fibre tip for effective coupling. The combination of two lenses, $\mathrm{L_2}$ and $\mathrm{L_3}$ in each arm, image the modes generated in the crystal plane onto the SLM planes. The lens after the SLM, $\mathrm{L_4}$ along with the aspheric lens within the fiber coupler, $\mathrm{L_{FC}}$, image the modes generated after phase-flattening at the SLM onto the fiber tip. The lenses are chosen such that the spatial mode sizes match the mode field diameter of the fiber coupling system. OAM projections in the even-odd basis are carried out with the help of identical spatial light modulators kept in the two arms. The holograms corresponding to superpositions of even orders and odd orders act as the counterparts of D and A projections in the HV basis.
\section{Results and Discussion}
While passing through the OAM sorting setup given in Fig. \ref{fig:OAMsortPOLent}, the down converted photons alternately choose between the even and odd ports depending upon their OAM value. Before making polarization entanglement measurements on these photons, the action of even-odd sorting within our setup needs to be verified.
\begin{figure}[h]
    \centering
    \includegraphics[width=\columnwidth]{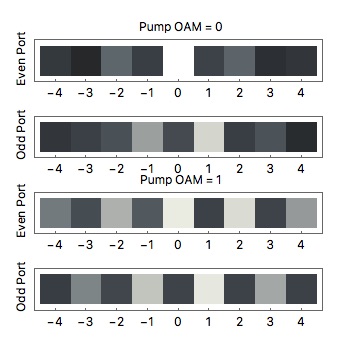}
    \caption{Verification of sorting of even and odd OAM states in the folded Mach-Zehnder sorter. The top two rows correspond to the outputs in the even and odd ports when pumped with a Gaussian ($l_p$ =0) and the bottom two rows correspond to that for a pump carrying $l_p$ =1 OAM.}
    \label{fig:sorted}
\end{figure}
The OAM state of the photons in the output ports of the sorter are measured using the standard technique involving phase-flattening through SLM and coupling to single mode fibers. The singles in each port corresponding to different OAM values are represented as a grayscale chart in Fig. \ref{fig:sorted}. The values are normalized with respect to the number photons in the Gaussian mode when pumped with a Gaussian beam, being the largest among all. When pumped with a Gaussian beam, photons are down converted in pairs of odd-odd or even-even pairs, following the conservation of OAM, and thus the photon pairs end up in the same port. It can be easily seen from the chart that photons carrying even and odd OAM values line up in the corresponding ports and their intensities are defined by the OAM spectrum of the SPDC output. For a pump carrying OAM, $l_p = 1$, the pairs are generated in even-odd pairs and they go to different ports. This is evident from how the corresponding intensity values are distributed between the two ports. For example, 0 in even port and 1 in odd port show similar intensity since they are generated together and so on. Moreover, the stark complementary behaviour in the intensity corresponding to even and odd ports for a partiular OAM value shows the effective sorting in the setup.

\begin{figure}[h]
\centering
    \includegraphics[width=\columnwidth]{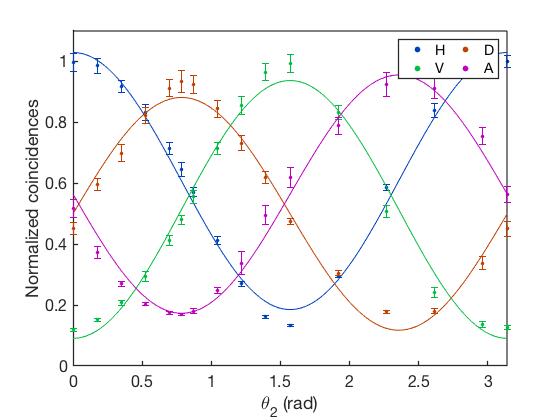}
    \caption{Polarization correlations corresponding to projections in the output ports of the even-odd sorted collinear SPDC output. Error bars indicate statistical uncertainty of one standard deviation.}
    \label{fig:polviscoll}
\end{figure}

Figure \ref{fig:polviscoll} shows the polarization correlations between the even and odd output ports of the sorter [Fig. \ref{fig:OAMsortPOLent}]. The indistinguishable photons are efficiently sorted under the label of their orbital angular momentum and polarization correlations are observed in both HV as well as DA basis.
The normalized coincidences are plotted along y-axis with the variation of $\theta_2$ for (blue) $\theta_1$ = 0$\degree$, (red) $\theta_1$ = 45$\degree$, (green) $\theta_1$ = 90$\degree$ and (purple) $\theta_1$ = 135$\degree$. The observed visibilities are 77.5$\pm$0.3\% (HV basis) and 71.6$\pm$0.3\% (DA basis).

The Bell parameter is estimated to be, S = 2.11 $\pm$ 0.03. It can be seen in the plot that the minima corresponding to different visibility profiles are not going completely to zero. This is due to the possible leakage of even OAM modes into the odd port (and vice versa). The reduced visibility can be understood as a manifestation of the imperfections in the sorting interferometer. Hence, with an improved interferometric sorter, it is possible to obtain near unity visibility.

\begin{figure}[h]
\centering
    \includegraphics[width=\columnwidth]{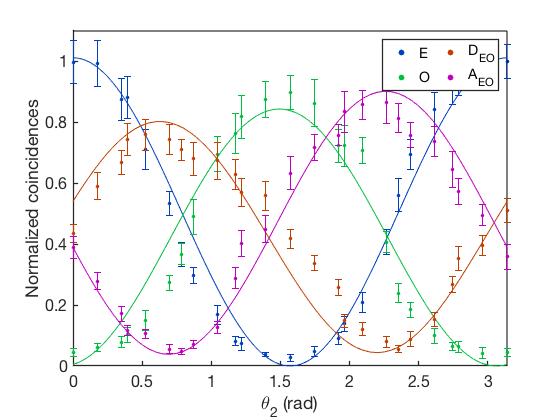}
    \caption{OAM correlations corresponding to two orthogonal polarization projections in the collinear SPDC output. Error bars indicate statistical uncertainty of one standard deviation.}
    \label{fig:oamviscoll}
\end{figure}

The OAM correlations between the H and V output ports of the polarizing beam splitter [Fig. \ref{fig:POLsortOAMent}] in the even-odd basis of OAM is given in Fig. \ref{fig:oamviscoll}. The indistinguishable photons are efficiently sorted under the label of their polarization and OAM visibility is observed in both EO as well as $\mathrm{D_{EO}A_{EO}}$ basis. The normalized coincidences are plotted along y-axis with the variation of $\theta_2$ for (blue) $\theta_1$ = 0$\degree$, (red) $\theta_1$ = 45$\degree$, (green) $\theta_1$ = 90$\degree$ and (purple) $\theta_1$ = 135$\degree$.
A calculation of visibility gives 92.7$\pm$0.3\% (EO basis) and 80.9$\pm$0.3\% ($\mathrm{D_{EO}A_{EO}}$ basis). The Bell parameter is estimated to be, S = 2.46 $\pm$ 0.08.

While one is required to alight multiple interferometers along with sorters to undertake a general set of projective measurements in the true linear and diagonal even-odd basis \cite{cbhanuevenodd}, we have carried out the measurements through OAM projections using SLM. The use of SLM however introduces efficiency constraints and limitations in exploting all the available OAM modes. Since our aim is to demonstrate the duality of entanglement in our setup, we have made the measurements in a basis defined by $l \in \{1,2\}$ where 1 and 2 represents the odd and even OAM states respectively. However, for practical applications, the projective measurements in the even-odd basis needs to be done as given in Ref. \cite{cbhanuevenodd} in order to explore all the available photons. The difference between the visibility of E and O curves in Fig. \ref{fig:oamviscoll} can be easily understood as the result of our choice of a reduced OAM basis consisting of only 1 and 2. Equal visibility could be achieved by doing the measurements in the full even-odd basis as mentioned above.
\section{Conclusions}
In this paper, we demonstrate the duality in entanglement of a collinear, indistinguishable pair of photons generated in a spontaneous parametric down conversion process. We show polarization entanglement for indistinguishable photons by sorting the photon OAM using a double Mach-Zehnder even-odd sorter. This method can increase the availability of entangled photons since we are not eliminating any photon from the generated output in contrast to the case of limiting them to two-dimensional OAM bases such as ($+l,-l$) or ($0,l$). All the down-converted photons are sorted using an even-odd sorter in order to observe the polarization entanglement of otherwise indistinguishable collinear photons. Similarly, we demonstrate OAM entanglement by sorting photons using a simple polarizing beam splitter and executing OAM projections on the photon pairs in the even-odd basis.

The entanglement studies of systems that display duality can give identical results in both the variables. This could evoke great interest in studying entanglement unaffected by the mutual interaction of particles. By further improving the efficiency of the sorter and incorporating all available OAM modes in even-odd projections, the entanglement measures estimated in the paper could be improved and could be shown to be more identical than obtained. In addition, duality assisted observation of entanglement can be used as a test for verification of indistinguishability of photons in quantum information processing. The measurement of entanglement in the complimentary variables could also reveal any change in the indistinguishability of photons over communication channels which could have arisen due to possible eavesdropping. The indistinguishable photons entangled in complementary variables may also find applications in distributed quantum sensing through phase estimation as well as remote entanglement generation in a quantum network.

\section*{Acknowledgement}
We are grateful to Dr. Ali Anwar and Mr. Pranav Tiwari for their valuable suggestions and discussion.

\bibliography{reference.bib}

\end{document}